\documentclass[english,12pt,aps,prd,a4paper,preprintnumbers,floatfix,nofootinbib,showpacs,superscriptaddress, notitlepage]{revtex4-1}
\pdfoutput=1

\usepackage{graphicx}
\usepackage{rotate}

\usepackage[utf8]{inputenc}

\usepackage{amsmath}
\usepackage{amssymb}
\usepackage{float}
\usepackage{comment}

\usepackage{soul}

\usepackage[usenames,dvipsnames]{color}
\usepackage[colorinlistoftodos]{todonotes}
\usepackage{amsmath}
\usepackage{amssymb}
\usepackage[colorlinks=true,citecolor=blue,urlcolor=blue, pdfborder={0 0 0}]{hyperref}
\usepackage[normalem]{ulem}
\newcommand {\ignore}[1]{}

\usepackage{makecell}
\definecolor{darkred}{rgb}{0.6,0,0}

\usepackage[english]{babel}
\usepackage{blindtext}

\definecolor{midnightblue}{RGB}{25,25,112}
\definecolor{brown}{rgb}{0.59, 0.29, 0.0}

\newcommand{\AddrIISER}{Department of Chemistry,\\
  Indian Institute of Science Education and Research (IISER) Bhopal \\
 Bhopal Bypass Road, Bhauri, Bhopal 462066 INDIA}

\begin{document}
\bibliographystyle{unsrt}   

\title{A near analytic solution of a stochastic immune response model considering variability in virus and T cell dynamics}
\author{Abhilasha Batra}
\affiliation{\AddrIISER}
\author{Rati Sharma}\email{rati@iiserb.ac.in}
\affiliation{\AddrIISER}

\begin{abstract}
\vspace{1cm}

Biological processes at the cellular level are stochastic in nature, and the immune response system is no different. Therefore, models that attempt to explain this system need to also incorporate noise or fluctuations that can account for the observed variability. In this work, a stochastic model of the immune response system is presented in terms of the dynamics of the T cells and the virus particles. Making use of the Green's function and the Wilemski-Fixman approximation, this model is then solved to obtain the analytical expression for the joint probability density function of these variables in the early and late stages of infection. This is then also used to calculate the average level of virus particles in the system. Upon comparing the theoretically predicted average virus levels to those of COVID-19 patients, it is hypothesized that the long lived dynamics that are characteristic of such viral infections are due to the long range correlations in the temporal fluctuations of the virions. This model therefore provides an insight into the effects of noise on viral dynamics.    

\end{abstract}

\maketitle

\section{Introduction}
The immune system of an organism provides the protection that it needs against foreign bodies such as microbes, viruses, parasites and more. The appearance of these foreign entities inside the body triggers the immune response, which is an agglomeration of cells, tissues and biochemical processes that function in conjunction with each other to protect the body \cite{PARKIN}. To stop the proliferation of such antigens, the immune system proliferates its own cells and shields the body from foreign attack, thereby enabling it to carry out its regular functions \cite{Perelson1997}.   
\\
The immune response in humans can be understood as a system with two levels of increasing complexity. These are the innate and the adaptive immunity. Innate immunity is the first line of defense and is non-specific. 
It has no immunologic memory as it cannot distinguish self from non-self \cite{CHAPLIN2010}. Therefore, after the pathogens manage to evade the innate immune system, the second line of defense, i.e., the adaptive immune response gets activated.
The adaptive/acquired immunity releases antigen-specific response and therefore provides a targeted defense against foreign particles. 
It works by retaining the copies of antibodies produced in the previous attack, therefore encoding the new memory for future use. 
This helps the immune system in launching a faster and targeted attack on foreign bodies in the future. The adaptive immunity is provided by a combined effort of two types of lymphocytes, which are, antigen-specific T cells (matured in thymus) and B cells (matured in the bone marrow) which divide into plasma to produce antibodies. 
T cells have surface specific antibody-like receptors that can recognize antigens inside the target cell of the host's body carrying the virus and directly destroys them  \cite{Marshall2018}.
\\
Since T cells are responsible for the directed attack on virus particles, a study of the dynamics between these two is pertinent. Theoretical studies of such systems are helpful in giving an insight into the complex dynamical phenomena involving their interaction. Therefore, several groups in the past have developed theoretical models to analyze and predict the virus and T cell dynamics in the system. 
These include models developed to look at the integrated immunological response to different viral infections such as Human Immunodeficiency Virus (HIV), Influenza virus, Zika virus and the virus that caused the ongoing pandemic, severe acute respiratory syndrome coronavirus 2 (SARS-CoV-2) \cite{Perelson2002,Boianelli2015,Best2018,Hernandez-Vargas2020}, among others. These theoretical studies generally use one of the two modelling approaches, viz. models without and with the immune response. The models without immune response generally incorporate kinetic interactions between healthy, susceptible and infected cells with the virus particles. Various standard versions of models using this approach, such as, ``Target Cell Limited model'' \cite{Best2018} and ``Target Cell Eclipse Phase model'' \cite{Baccam7590,Wang2020} have been studied. On the other hand, the models that include the immune response deal with interactions between immune cells (T cells) and viral particles \cite{Hernandez-Vargas2020,Almocera2018}. The immune response model was also developed for the Influenza A virus, where, the dynamics of cytotoxic T cells and virus population were coupled through a set of coupled ODEs \cite{Boianelli2015}. 
%
\\
%
All of these models discussed above are deterministic. 
However, biological processes at the cellular level, including the immune response systems are stochastic in nature \cite{lzaro10830}. Therefore, one must take care to incorporate fluctuations/noise in these systems. In fact, an analysis of experimental data has shown that the presence of random noise in gene expression leads to increased variability in viral (HIV) gene products such as RNA, which contributes to the replication of viral material and the latency period \cite{Singh2010}. Stochasticity has also been reported in the division and death time of lymphocytes \cite{Hawkins2007}. 
\\
Taking this variability and population heterogeneity into consideration, a few stochastic immune response models have been developed in the recent past \cite{Dalal2008, Wang2019a,Fatehi2018}. Dalal et al. in their work introduced stochasticity in the deterministic model of immune response in HIV infection by parameter perturbation \cite{Dalal2008}. Wang et al. showed that the stability of the stochastic dynamical system of HIV infection is different when fluctuations are introduced in terms of the Gaussian colored noise in contrast to the Gaussian white noise \cite{Wang2019a}. Another recent stochastic model of T cell dynamics was used to explain the bistability and crossover dynamics of the immune response \cite{Roy2020}. These stochastic studies indicate that modeling fluctuations into the system can give a more accurate picture of the immune response dynamics.  
\\
In this work, we aim to look at the stochastic nature of the immune response with special focus on the SARS-CoV-2 virus for the ongoing global COVID-19 pandemic. To do this, we model the immune response dynamics in viral infections by incorporating stochastic fluctuations in terms of the Gaussian white noise and the fractional Gaussian noise to the set of coupled ODEs of T cells and virus particles. Our stochastic immune response model provides a near analytic solution for the time dependent joint probability distribution of T cells and virus, which we then use to determine the average number of virus particles in the system. 
%
%
Considering the ongoing pandemic, we then also compare our results to the experimental SARS-CoV-2 viral infection data from Germany \cite{BOHMER2020,Wolfel2020}. The long mean incubation period, which is approximated to be 5-6 days \cite{ANDERSON2020c,Lauer} makes it important to analyse the viral dynamics of SARS-CoV-2 using the stochastic immune response model. Although, in this work, we carry out the numerical analysis for this particular virus, similar analysis can also be applied to other viral dynamics.
\\
This paper is organized as follows: In Section II, we formulate the stochastic immune response model using a set of coupled stochastic differential equations (SDEs). Section III provides the derivation of the temporal evolution of the joint probability distribution (Fokker-Planck equation) of T cells and virus particles from the set of coupled SDEs. Part A of the results section includes a near analytic solution of Fokker-Planck equation in different time regimes. The analysis carried out here is generic and is applicable to viral infections in general. Part B has two subsections to show the numerical analysis of the stochastic quantities of immune response in SARS-CoV-2 infection. We finally summarize our results in Section V.

\section{Stochastic immune response model}

The simplest model of immune response dynamics needs to account for interactions between the immune cells and the virus. The immune response itself is activated when cells are under attack by foreign bodies such as viruses. This activation is manifested through an increased proliferation of the immune cells. Therefore, there needs to be a coupled interaction between the immune cells and the virus. Since the proliferation and death rates of both these entities are intrinsically stochastic \cite{Hawkins2007, Singh2010a}, 
any realistic model needs to account for this inherent variability as well. The inherent intrinsic variability can be accounted for in the model via the Gaussian white noise (GWN), which is delta correlated. The GWN is a fast decaying noise, where successive fluctuations are not correlated. We model the inherent noise in T cells as a GWN. Viruses though are known to show population level fluctuations between active and latent states \cite{Singh2010}. This is a manifestation of fluctuations in the gene expression, which in turn leads to variability in the gene products. We therefore model the viral dynamics to evolve under the action of the fractional Gaussian noise (fGn), which is a type of colored noise. Specifically, the dynamical interaction between T cells and the virus particles can then be written as a set of coupled stochastic differential equations (SDEs), given by 
\begin{eqnarray}\label{xdyn}
    \Dot{x}(t) = \beta \big(v(t)\big) x(t) - \gamma x(t)+\theta (t) 
\end{eqnarray}
\begin{equation}\label{vdyn}
     \int_{0}^{t} dt' K(t-t') \Dot{v}(t') = pv(t)-cv(t)+\xi(t)
\end{equation}
%

\begin{figure}[hbpt]
\includegraphics[width=\linewidth]{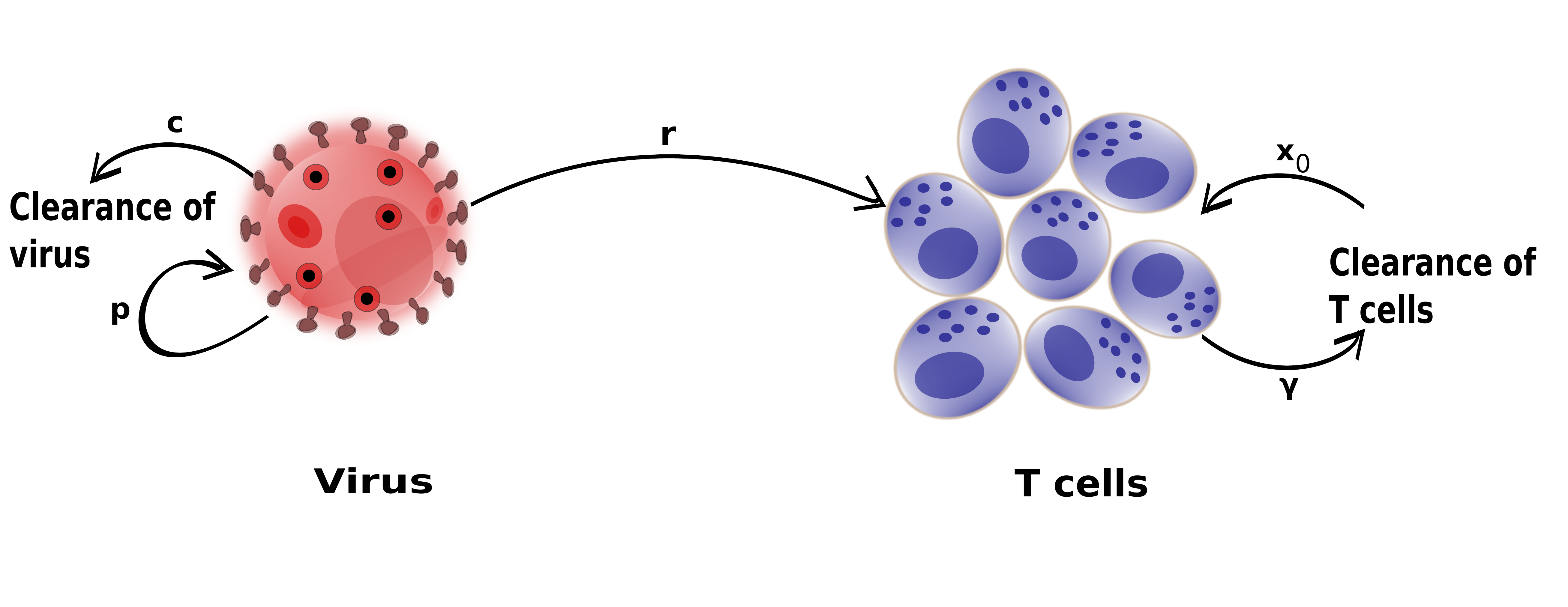}
\caption{\textbf{Schematic representation of stochastic immune response model}: $r$ is the proliferation rate of T cells, $p$ is the replication rate of virus particles, $x_0$ is the production rate of basal level T cells,  $c$ and $\gamma$ are clearance rates of the virus particles and the T cells respectively. [These are representative images, not to scale]}
\label{fig:Schematic_diagram}
\centering
\end{figure}

Here, $x(t)$ and $v(t)$ represent the concentration levels of T cells and virus particles in the body which are cleared at rates $\gamma$ and $c$ respectively. $p$ is the replication rate of virus particles. $\beta\big(v(t)\big)$ is a function of virions concentration, which incorporates the effect of virus levels in T cell dynamics.  $\beta\big(v(t)\big)$ is a positive odd integer power law function that accounts for the dependency between T cells and virus levels. This term therefore couples the viral dynamics to T-cell dynamics. In general, one can consider $\beta\big(v(t)\big) = r v^m$ to account for the fact that higher the value of $m$, faster is the rate of increase of $\beta\big(v(t)\big)$. Earlier immune response models \cite{Boianelli2015} also indicate the same, i.e., increase in virus levels lead to proliferation of T cells with a rate $r$. In our study, we have set the value of $m$ to be 1. A schematic representation of the stochastic immune response model is shown in Fig. \ref{fig:Schematic_diagram}.
\\
The terms $\theta (t)$ and $\xi (t)$ represent GWN and fGn with noise strengths $a$ and $\lambda$ respectively. These account for the variability in the system and have the following statistical properties.

\begin{align}\label{noise_prop}
\begin{split}
\left< \theta(t) \right> &= 0\\
\left< \xi(t) \right> &= 0\\
\left< \theta (t) \theta (t')\right> &= a \delta(t-t')\\
\left<\xi(t)\xi(t')\right> &= \lambda K(|t-t'|)\\
\end{split}
\end{align}

Here, the angular brackets represent an average over all realizations of the noise.
As specified in Eq\eqref{noise_prop}, both the noise terms, $\theta(t)$  and $\xi(t)$, have zero mean. $\theta (t)$ represents fluctuations of a Markov process which is delta correlated, i.e., the fluctuation at any time $t$ is uncorrelated with the previous time $t'$, while $\xi(t)$ represents correlated fluctuations characteristic of fractional Gaussian noise (fGn).
fGn is a Non-Markovian process, which is temporally correlated by a memory kernel $K(|t-t'|)$, which has the following form \cite{Kou2004,Min2005,Mandelbrot,Lim2002,Fa2005} 
\begin{eqnarray}\label{memoryfn}
 K(|t-t'|) &= 2H(2H-1)|t-t'|^{2H-2}
\end{eqnarray}
Here $H$ is the Hurst index such that $1/2 \le H < 1$. The value $H=1/2$ represents the delta correlated limit of the memory kernel, whereas away from this value, the fluctuations become more correlated. 
The strength of noise for GWN and fGn  \textit{i.e.} $a$ and $\lambda$ respectively, specifies the deviation of noise from its mean. 

\section{Transformation to the Fokker-Planck Equation}
The advantage of representing a system through a stochastic model is the possibility of obtaining a multivariate probability distribution function from the set of SDEs, Eqs. \eqref{xdyn} and \eqref{vdyn}. This distribution function can then be used to obtain the average values of the relevant variables, in this case, the number of virions inside the host's body at a given time. 
\\
To begin with, we first define the distribution function $P(x,v,t)$ as the probability density of finding $x$ T cells and $v$ virions at a particular time $t$. This can be written as 
\begin{equation}\label{prob_general}
    P(x,v,t) = \left<\delta (x-x(t)) \delta(v-v(t))\right>
\end{equation} 
where $x(t)$ and $v(t)$ are functionals of the noise $\theta(t)$ and $\xi(t)$ respectively and the angular brackets represent an average over all realizations of the noise. Now, substituting the solutions of the differential equations, Eqs\eqref{xdyn} and \eqref{vdyn} into Eq \eqref{prob_general} and making use of the noise properties (Eq\eqref{noise_prop}), one can obtain the time evolution of the probability density function $P(x,v,t)$. This equation, known as the Fokker-Planck equation, is given by

\begin{equation}\label{FPE_oper}
    \frac{\partial}{\partial t}P(x,v,t) = -\beta\big(v(t)\big)P(x,v,t)- \mathbf{L} P(x,v,t)
\end{equation}
where the operator $\mathbf{L}$ is defined as 
\begin{equation}\label{operator}
        \mathbf{L} = \beta\big(v(t)\big) x \frac{\partial}{\partial x} - \gamma \frac{\partial}{\partial x} x -\frac{1 }{2}a\frac{\partial^2}{\partial x^2}- \eta(t)\frac{\partial}{\partial v} v -\frac{\lambda}{|(c-p)|}\eta(t)\frac{\partial^2}{\partial v^2}
\end{equation}

Here, $\eta(t)$ is a time-dependent function defined as $\eta(t) \equiv -\mathcal{\Dot{X}}(t)/{\mathcal{X}(t)}$, where $\mathcal{{X}}(t) = \mathbf{E}_{2-2H}(-(t/\tau)^{2-2H})$ and $\tau = \left(\frac{\Gamma(2H+1)}{|(c-p)|}\right)^{1/(2-2H)}$. $\mathbf{E}_{\alpha,\beta}{(z)}$ is a Mittag-Leffler function of the form $\sum^{\infty}_{n=0} z^n/\Gamma(\alpha n + \beta)$, where $\Gamma{(\alpha n + \beta)}$ is a gamma function.\\

The details of this transformation (from Eqs\eqref{xdyn} and \eqref{vdyn} to Eq\eqref{FPE_oper}) are shown in Appendix A.
The exact joint probability distribution function, $P(x,v,t)$, can be obtained from the time dependent solution of the Fokker-Planck Equation (Eq\eqref{FPE_oper}), which will then be useful in studying the various statistics of T cells and virus particles.\\

\section{Results}
\subsection{Solution of the Fokker-Planck Equation: Obtaining $\mathbf{P(x,v,t)}$}
The Fokker-Planck Equation, Eq\eqref{FPE_oper} can be represented through an equivalent form by making use of the Green's function. The system evolves with time from its equilibrium state and therefore the formal solution of Eq\eqref{FPE_oper} is given by \cite{Wilemski2001}
\begin{align}\label{WF app}
    \begin{split}
        P(x,v,t) = {P_{eq}}(x,v) - \int_{0}^{\infty}dx'\int_{0}^{\infty}dv' \int_{0}^{t}dt'G(x,v,t-t'|x',v') \beta(v')P(x',v',t')
    \end{split}
\end{align}
Here the Green's function, $G(x,v,t-t'|x_0,v_0)$, is the time dependent conditional probability of finding the system in the state $(x,v)$ at time $t$ given that it was in the state $\left(x_0, v_0\right)$ at time $t=0$. The detailed derivation of the Green's function using the operator $\mathbf{L}$ is provided in Appendix B. After some lengthy algebra, the Green's function (propagator) is given by
\begin{equation}\label{Greens function}
\begin{split}
G(x,v,t|x_0,v_0,0) = \frac{1}{2 \pi}\frac{1}{\sqrt{\frac{a}{2 \gamma}\frac{\lambda}{ |(c-p)|}{(1-e^{-2 \gamma t})}{(1 - \mathcal{X}^2(t))}}}\exp\bigg( \frac{- 1}{2}\bigg( \frac{1}{\frac{a}{2 \gamma}} \frac{(x - x_0 e^{-\gamma t})^2}{ (1-e^{-2 \gamma t})}\\+\frac{1}{\frac{\lambda}{ |(c-p)|} }\frac{(v - v_0 \mathcal{X} (t))^2}{ (1-\mathcal{X}^2 (t))}\bigg) \bigg)
\end{split}
\end{equation}
 
This propagator satisfies the condition that when $t\rightarrow \infty$, the time dependent conditional probability density , $G(x,v,t-t'|x',v')= P_{eq}(x,v)$. Therefore,
\begin{equation}\label{eqm prob}
   {P_{eq}}(x,v) = \frac{1}{2 \pi}\frac{1}{\sqrt{\frac{a}{2 \gamma} \frac{\lambda}{ |(c-p)|}}} \exp{\left ( \frac{-1}{2}\left( \frac{x^2}{(\frac{a}{2 \gamma})} + \frac{v^2}{\frac{\lambda}{|(c-p)|}} \right)\right)}
\end{equation}
Eq\eqref{WF app} can provide the implicit solution for $P(x,v,t)$, but to determine the analytic expression for the time dependent joint probability distribution function, we use the closure scheme introduced by Wilemski and Fixman \cite{Wilemski1974e,Wilemski1974b}. The Wilemski-Fixman (WF) approximation was originally developed to estimate the rate of the reaction between reactive groups at either ends of a polymer chain \cite{Wilemski1974b}. This was later extended to the systems of catalytic
bimolecular reactions \cite{Benichou2005}, non-exponential DNA escape kinetics \cite{Chatterjee2010a}, dynamic disorder in chain unfolding \cite{Chatterjee2011a} and chain closure in entangled polymer systems \cite{Bhattacharyya2012}, among others. In the original work by Wilemski and Fixman \cite{Wilemski1974e}, a sink term was introduced to provide a valid ``closure approximation'' for the  diffusion equation in the many-particle system of polymer reactions. Using this approximation, the solution of Eq\eqref{WF app} can be replaced by an approximate expression which involves the product of two terms - (i) the equilibrium probability density $P_{eq}(x,v)$, that is time independent and corresponds to the situation when T cells and virus levels in the system are independent of each other and (ii) a self-consistently determined time-dependent term which evolves with time from the equilibrium distribution ($P_{eq}(x,v)$) as a consequence of the sink term, i.e., $\beta\big(v(t)\big)$ in the case of our stochastic immune response model.

Therefore using this approximation, the probability distribution function $P(x,v,t)$ can be defined by the introduction of two functions $w(t)$ and $\Bar{w}$ such that,
\begin{equation}\label{probab}
        P(x,v,t) = P_{eq}(x,v)\frac{ w(t)}{\Bar{w}}
\end{equation}
where, 
\begin{equation}\label{wt}
 w(t) = \int_{0}^{\infty} \int_{0}^{\infty} dx dv \beta(v)P(x,v,t) \quad \text{and} \quad
    \Bar{w} = \int_{0}^{\infty} \int_{0}^{\infty} dx dv \beta(v)P_{eq}(x,v)
\end{equation}
Multiplying Eq\eqref{WF app} by $\beta (v)$, integrating over $x$ and $v$ and then substituting Eq\eqref{probab} into it provides the expression for $w(t)$ such that
\begin{eqnarray}\label{wt final}
        w(t) = \Bar{w} -\int_{0}^{t} dt'C(t-t')w(t')/\Bar{w}
\end{eqnarray}
where
\begin{eqnarray}\label{C(t)}
        C(t-t') = \int_{0}^{\infty} dx'\int_{0}^{\infty} dx\int_{0}^{\infty} dv'\int_{0}^{\infty} dv \beta(v) G(x,v,t-t'|x',v')\beta(v')P_{eq}(x',v')
\end{eqnarray}
Following the above steps, one can see that $w(t)$ is required in the calculation of the probability distribution for the immune response model, which in turn requires the evaluation of $C(t)$. Computation of $C(t)$ is carried out by substituting Eq\eqref{Greens function}, Eq\eqref{eqm prob} and the expression for $\beta(v)$ into Eq [\ref{C(t)}] and then carrying out the integration. This gives
\begin{equation}\label{C(t) final}
    \begin{split}
        C(t) = \left(\frac{\pi}{4}\right)^2 r^2 \frac{\lambda}{ |(c-p)|} \left(2 \sqrt{1-\mathcal{X}^2 (t)}+ \mathcal{X}(t) \left(\pi + 2 \textrm{ArcTan} \left(\frac{\mathcal{X}(t)}{\sqrt{1-\mathcal{X}^2 (t)}}\right)\right)\right)\\ \left(\pi + 2 \textrm{ArcTan}\left(\frac{e^{-\gamma t}}{\sqrt{1-e^{-2 \gamma t}}}\right)\right)
    \end{split}
\end{equation}
$w(t)$ can simply be obtained by using the method of Laplace transforms. This requires the Laplace transform of $C(t)$ as well. Determining the simple algebraic form of the Laplace transform of Eq\eqref{C(t) final} is non-trivial due to the presence of $\mathcal{X}(t)$ i.e. the Mittag-Leffler function. However, viral dynamics are particularly interesting during the early and late stages of infection. Viral populations reach their peaks in the early stage of infection and are especially long lived and decay gradually during the late stages of infection \cite{Wolfel2020}. Therefore, here, we are primarily interested in viral dynamics, and in turn,  $\mathcal{X}(t)$ in two different time regimes i.e. at short and long times.
\subsubsection{Short time regime}
In the case of short times ($t/\tau<<1$ and $\gamma t<<1$), the Mittag-Leffler function is approximated to $\mathcal{X}(t) = 1-a_1 t^b+O(t^{2b})$, where $b = 2-2H$, $a_1 = 1/\tau^b \Gamma (3-2H)$. The functions $\; \textrm{ArcTan} \left(\frac{\mathcal{X}(t)}{\sqrt{1-\mathcal{X}^2 (t)}}\right) \approx \frac{\pi}{2}-\left(1-a_1 t^b\right)\sqrt{2a_1 t^{b}}$ and $\textrm{ArcTan} \left(\frac{e^{-\gamma t}}{\sqrt{1- e^{-2\gamma t}}}\right) \approx \frac{\pi}{2} - (1-\gamma t) \sqrt{2 \gamma t}$,  for small value of $\gamma$ this approximates to $ \frac{\pi}{2}$.
Ignoring higher order terms of $t$, the expression obtained for $C(t)$ in the short time regime is,
\begin{eqnarray}\label{C(t) st}
    C(t) = \left(\frac{\pi^2}{2}\right)^2 r^2 \vartheta  \left(1 - a_1 t^b \right)
\end{eqnarray}
where $\vartheta = {\lambda/|(c-p)|}$\\
Substituting the Laplace transform of Eq\eqref{C(t) st} into the Laplace Transform of Eq\eqref{wt final}, i.e., into  $ w(s) = \Bar{w} / \left({s\left(1+\frac{C(s)}{\Bar{w}}\right)}\right)$, one can obtain the complete expression for $w(s)$. This is given by 
\begin{equation}\label{w(s) st}
 w(s) = \frac{4\Bar{w} s^b}{(r^2 \pi^4 \vartheta s^b/ \Bar{w} ) + 4 s^{1+b}- (r^{2} \pi^{4} \lambda/(\Bar{w} \Gamma(3-b)))}
\end{equation}
The series expansion of the above expression gives
\begin{equation}\label{w(s) series st}
w(s) = \Bar{w} {\sum}_{k=0}^{\infty} (-1)^k \left(\frac{r^2 \vartheta \pi^4}{4 \Bar{w}}\right)^k \frac{s^{b(k+1)}}{\left(s^{1+b}-A\right) ^{k+1}}
\end{equation}
where $A = \frac{(r \pi^{2})^2}{4 \Bar{w}} \frac{\lambda}{\Gamma(3-b)}$. The inverse Laplace transform of Eq\eqref{w(s) series st} provides the expression for $w(t)$ in the short time regime. This is given by
\begin{eqnarray}\label{w(t) st kth}
    w(t) = \Bar{w} {\sum}_{k=0}^{\infty} \frac{(-1)^k}{k!} \left(\frac{r^2 \vartheta \pi^4}{4 \Bar{w}}\right)^k t^k \mathbf{E}^{(k)}_{1+b,1-bk}\left(A t^{1+b} \right)
\end{eqnarray}
where $\mathbf{E}^{k}_{\alpha,\beta}{(z)}= \frac{d^k \mathbf{E}_{\alpha,\beta}{(z)}}{d z^k}$ is the $k^{th}$ derivative of the Mittag-Leffler function with respect to its argument.
The series expansion of the Mittag-Leffler function \cite{Vinales2006} is given by $\mathbf{E}_{\alpha,\beta}{(z)}={\sum}_{n=0}^{\infty} \frac{z^n}{\Gamma{(\alpha n + \beta)}}$
whose $k^{th}$ derivative turns out to be 
\begin{equation}\label{Ek derivative}
\mathbf{E}^{(k)}_{1+b,1-bk}\left(A t^{1+b}\right) = {\sum}_{n=0}^{\infty}\frac{(n+k)!}{n!}\frac{(A t^{1+b})^n}{\Gamma(n(1+b)+k+1)}
\end{equation}
Substitution of Eq\eqref{Ek derivative} in Eq\eqref{w(t) st kth} results in the expression for $w(t)$. For short times (i.e., taking into account only the $k=0$ term as in \cite{Vinales2006}), the final expression is 
\begin{equation}\label{w(t) final st}
    w(t) = \Bar{w} \mathbf{E}_{1+b,1}\left(A t^{1+b}\right)
\end{equation}
where $\Bar{w} = \frac{1}{2 \sqrt{2 \pi}}{r\sqrt{\vartheta}}$. 
Substituting the above expression into Eq\eqref{probab} gives the joint probability density expression, $P(x,v,t)$, in the short time regime, which is\\
\begin{equation}\label{prob st}
    \begin{split}
        P(x,v,t) = \frac{1}{2 \pi \sqrt{\Im \vartheta}} \exp\left(\frac{-1}{2}\left(\frac{x^2}{\Im}+ \frac{v^2}{\vartheta}\right)\right)\mathbf{E}_{1+b,1}\left(\frac{\pi^{9/2}}{\sqrt 2}\frac{r}{\Gamma(3-b)}\sqrt{\lambda}\sqrt{|(c-p)|} t^{1+b}  \right)
    \end{split}
\end{equation}
where $\Im = {a}/{2 \gamma}$ and $\vartheta = {\lambda}/{|(c-p)|}$. The consequences of the dynamics in the short time regime will be discussed in Part B of this section.\\
\subsubsection{Long time regime}

Another regime of interest is the viral dynamics in the long time regime. Therefore, for large values of $t$, the Mittag-Leffler function is approximated to $\mathcal{X}(t) \approx a_2 t^{-b}$, where $a_2 = \tau^b / \Gamma{(2H-1)}$ and $b= 2-2H$ \cite{Chatterjee2010a}.
Further, ignoring higher order terms and using the property that $\textrm{ArcTan}(f(t))\approx f(t)$ when $f(t) << 1$, we can approximate the functions $\textrm{ArcTan} \left(\frac{\mathcal{X}(t)}{\sqrt{1-\mathcal{X}^2 (t)}}\right) \approx a_2 t^{-b}$ and 
   $\textrm{ArcTan} \left(\frac{e^{-\gamma t}}{\sqrt{1- e^{-2\gamma t}}}\right) \approx 0$. After applying these approximations, the closed form expression for $C(t)$ in the long time regime is given by
\begin{equation}\label{C(t) lt}
    C(t) = \left(\frac{\pi}{4}\right)^2 r^2 \vartheta \left(2\pi+\pi^2 a_2 t^{-b}\right)
\end{equation}
where $\vartheta = {\lambda/|(c-p)|}$\\
Substituting the Laplace transform of Eq\eqref{C(t) lt} into the Laplace Transform of Eq\eqref{wt final}, i.e., into  $ w(s) = \Bar{w} / \left({s\left(1+\frac{C(s)}{\Bar{w}}\right)}\right)$, we get 
\begin{equation}\label{w(s) lt}
    w(s) = \frac{16 \Bar{w}^2}{2 r^2 \vartheta\pi^3 \left(1 + \frac{16 \Bar{w} s}{2\pi^3 r^2 \vartheta} + \frac{\pi}{2}\frac{\Gamma(3-b)}{|(c-p)|} s^b \right)}
\end{equation}
The series expansion of Eq\eqref{w(s) lt} is
\begin{equation}\label{w(s) series}
    w(s) = \Bar{w}{\sum}_{k=0}^{\infty} \frac{(-1)^k}{k!} \left(\frac{2 \pi^3 r^2 \vartheta }{16 \Bar{w}}\right)^k k! \frac{s^{-b(k+1)}}{\left(s^{(1-b)}+A_2\right)^{k+1}}
\end{equation}
where $A_2 = \left(\frac{r \pi^{2}}{4}\right)^2 \frac{\Gamma{(3-b)}}{\Bar{w}}\frac{\lambda}{|(c-p)|^2}$. 
The inverse Laplace transform of Eq\eqref{w(s) series} provides
\begin{equation}\label{w(s) lt fn}
    w(t) = \Bar{w}{\sum}_{k=0}^{\infty} \frac{(-1)^k}{k!} \left(\frac{2 \pi^3 r^2 \vartheta }{16 \Bar{w}}\right)^k t^k \mathbf{E}^k_{1-b, 1+bk}\left(- A_2 t^{1-b}\right)
\end{equation}
Making use of the asymptotic form of the Mittag-Leffler function for larger values of its argument i.e. $\mathbf{E}_{\alpha,\beta}{(z)} = \frac{1}{z \Gamma{(\beta - \alpha)}}$, the $k^{th}$ derivative turns out to be 
\begin{equation}
    \mathbf{E}^k_{1-b, 1+bk}(- A_2 t^{1-b}) = \frac{k!}{\Gamma(b(1+k))} \left( A_2 t^{1-b} \right)^{-(k+1)}
\end{equation}



Upon substituting the above expression into Eq\eqref{w(s) lt}, the final expression for $w(t)$ in the long time regime is given by
\begin{equation}\label{w(t) lt final}
    w(t) = \frac{16 \Bar{w}^2}{a_2 r^2 \vartheta \pi^4 \Gamma{(1-b)} t^{1-b}} \mathbf{E}_{b,b}\left(\frac{-2 t^b}{\pi}\frac{|(c-p)|}{\Gamma(3-b)}\right)
\end{equation}
where $\Bar{w} = \frac{1}{2 \sqrt{2 \pi}}{r\sqrt{\vartheta}}$ and $\mathbf{E}_{b,b}$ is the Mittag-Leffler function.
The joint probability distribution function, $P(x,v,t)$ in the long time regime is then given by
\begin{equation}\label{prob lt}
    \begin{split}
       P(x,v,t) = \frac{8}{\pi^4}\frac{1}{(2\pi)^{3/2}r a_2 \sqrt{\Im \vartheta} \Gamma{(1-b)}} t^{b-1}\exp\left(\frac{-1}{2}\left(\frac{x^2}{\Im}+\frac{v^2}{\vartheta}\right)\right)\\ \mathbf{E}_{b,b}\left(\frac{-2 t^b}{\pi}\frac{|(c-p)|}{\Gamma(3-b)}\right)
    \end{split}
\end{equation}
where $\Im = {a}/{2 \gamma}$ and $\vartheta = {\lambda}/{|(c-p)|}$. The consequences of the dynamics in the long time regime will be discussed in Part B of this section. Equations [\ref{prob st}] and [\ref{prob lt}] are the main results of this work, which we further explore in the next section.\\

\subsection{Numerical implementations of stochastic immune response model}
The analytical results presented in the previous section, i.e., Eqs\eqref{prob st} and \eqref{prob lt}, provide the joint probability density function $P(x,v,t)$ for T cells and virus particles in different time regimes, namely, the early (short times) and late (long times) stages of infection. Following this calculation, we carry out its numerical analysis to study the temporal evolution of the time dependent joint probability distribution function. Our analysis applies in general to typical virus and immune response interactions, but, because the ongoing pandemic due to COVID-19 demands special attention, we analyze our results in the context of the immune response system for T cells and the  
SARS-CoV-2 virus, a pathogenic RNA virus with a lipid envelope. 
SARS-CoV-2 has long mean incubation period, which is approximated to be 5-6 days \cite{ANDERSON2020c}. The latency period i.e. the time lag between the time of infection and the onset of the initial symptoms, makes it pertinent to analyse the viral dynamics of SARS-CoV-2 using the stochastic immune response model. To gain an insight into the applicability of our model to the real world data, we have carried out a quantitative analysis by comparing our results to those obtained from the clinical data of COVID patients \cite{BOHMER2020,Wolfel2020}.\\
\begin{figure}
    \centering
    \includegraphics[width=\linewidth]{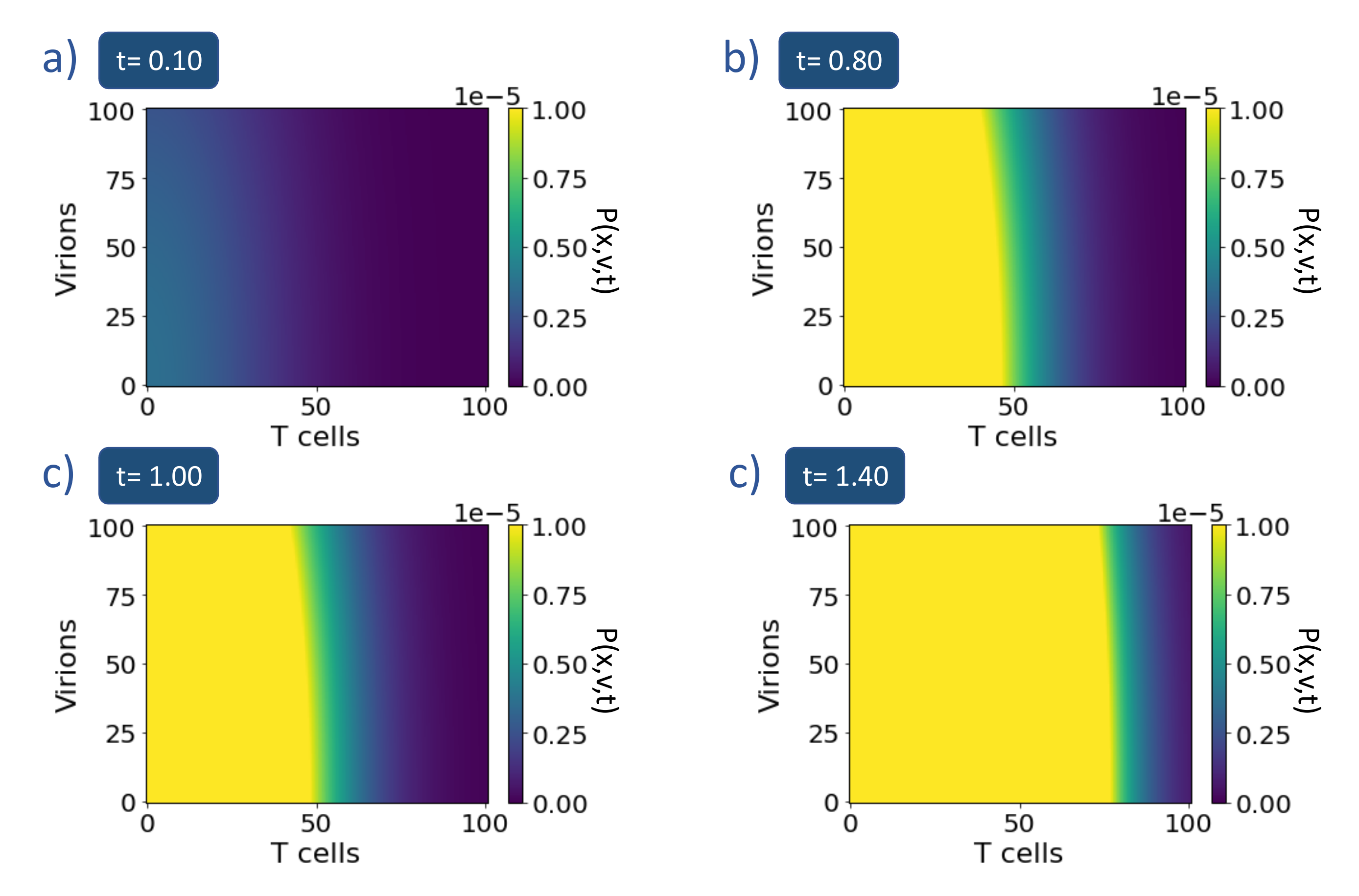}
    \caption{\textbf{Temporal evolution of joint probability distribution of T cells and virions during the early infection period, (short time regime) for $\mathbf{H=0.55}$:} a) $P(x,v)$ at 0.1 days. b) $P(x,v)$ at 0.8 days. c) $P(x,v)$ at 1 day. d) $P(x,v)$ at 1.4 days from the start of infection.}
    \label{fig2}
\end{figure}
\begin{figure}
    \centering
    \includegraphics[width=\linewidth]{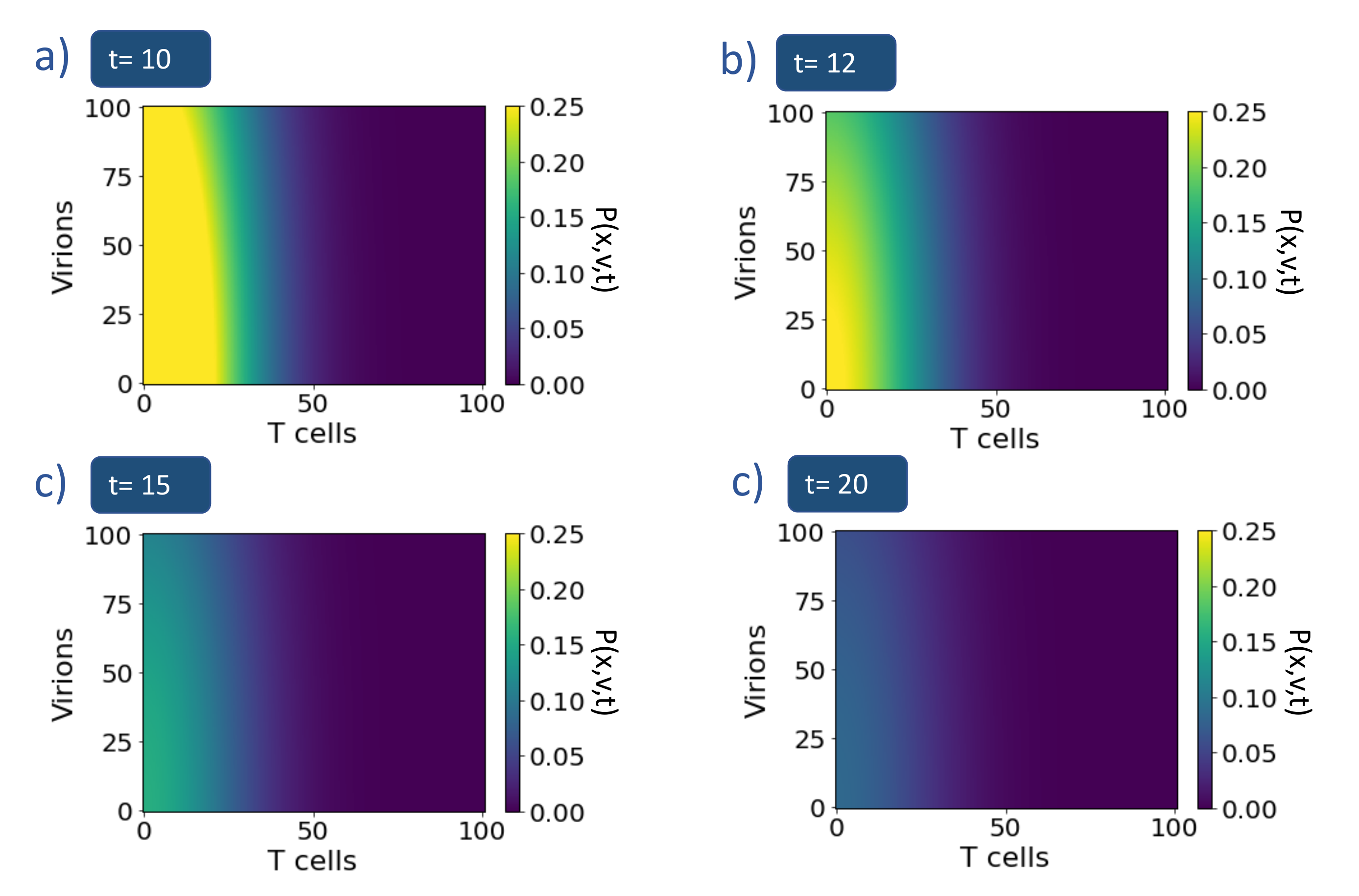}
    \caption{\textbf{Temporal evolution of joint probability distribution of T cells and virions during the late stage of infection, (the long time regime) for $\mathbf{H=0.55}$:} a) $P(x,v)$ after 10 days. b) $P(x,v)$ after 12 days. c) $P(x,v)$ after 15 days. d) $P(x,v)$ after 20 days of infection.}
    \label{fig3}
\end{figure}
\subsubsection{Temporal evolution of the joint probability distribution function}
The time dependent joint probability distribution expression for T cells and virus particles in the short and the long time regimes are given by Eqs\eqref{prob st} and \eqref{prob lt} respectively. In this work, we have studied the evolution of probability distribution function in two different time regimes: (a) from the time when SARS-CoV-2 enters the host's body to start the infection and (b) during the extinction of SARS-CoV-2 from a patients' body. The parameters required for this study have been determined from the earlier research studies \cite{Wang2020,Hernandez-Vargas2020} which have looked at the effect of SARS-CoV-2 on the basis of experimental data of viral load within the patients' bodies. The strength of the noise parameters, i.e., GWN in T cells and fGn in virus particles are $a=0.3$ and $\lambda = 0.1$ respectively. These are fit parameters that are variable. The clearance rate of T cells, $\gamma$ has been estimated on the basis of the half life of T cells, which is approximated to be 4 - 34 days \cite{McDonagh1995}, therefore in these calculations we set $\gamma$ to $0.1$ day$^{-1}$. Clearance rate of virus particles ($c$), proliferation rate of T cells ($r$), and replication rate of virus particles ($p$) are set to the mean values of these parameters for different patients' in different time regimes. These are listed in Table \ref{tab:tab1}.
Fig. \ref{fig2} shows the temporal evolution of the distribution function in the initial period of infection, approximately up to 2 days. Fig. \ref{fig3}, on the other hand provides an insight into the decrease in the joint probability distribution of T cells and virus particles at larger times i.e. during virus extinction and when T cell levels drop to basal values. These results show that the joint probability of T cells and virus particles increases with time at the start of infection and shows the decrease in probability distribution when virus levels attain a very low value (virus extinction period) within the host's body. 

\begin{table}
\begin{tabular}{ccccc}
    \Xhline{4\arrayrulewidth}
    Patients & Regime & $c$ $(day^{-1})$ & $p$ $(day^{-1})$ & $r$ (ml/cells/day)  \\
    \Xhline{4\arrayrulewidth}
    A & Short time&6.89 & 7.87 & 0.30\\
      & Long time & 7.50 & 4.20 & 0.33\\
    \hline
     B & Short time&5.32 & 6.81 & 0.33\\
      & Long time & 7.87& 4.37& 0.40\\
    \hline
    C & Short time&7.13 & 8.53 & 0.31\\
      & Long time & 7.67 & 5.19 & 0.39\\
    \hline
    D & Short time&4.93 & 5.54 & 0.34\\
      & Long time & 6.92 & 5.49& 0.26\\
    \hline
    E & Short time&4.98 & 5.33 & 0.31\\
      & Long time & 6.92 & 5.31 & 0.26\\
    \hline
    F & Short time&5.27 & 7.90 & 0.11\\
      & Long time & 7.04 & 4.22 & 0.20\\
    \hline 
    G & Short time&6.74 &7.62 & 0.26\\
      & Long time & 11.71 & 6.23 & 0.27\\
    \hline
    H & Short time&6.22 & 9.73 & 0.35\\
      & Long time & 15.07 & 9.12 & 0.39\\
    \Xhline{2\arrayrulewidth}
     & Short time & [4.93-7.13]  & [5.33-9.73]& [0.11-0.35]\\
     & Median&  5.77 &  7.74 & 0.31\\
    \Xhline{2\arrayrulewidth}
    &  Long time & [6.92-15.07] & [4.20-9.12] & [0.20-0.40]\\
      &Median  &  7.58  &  5.25 & 0.30\\
    \Xhline{4\arrayrulewidth}
\end{tabular}
\caption{Numerical values of rate parameters used in the quantitative analysis of the stochastic immune response model for SARS-CoV-2 virus. $c$ and $p$ are respectively the clearance rate and the replication rate of virus particles. $r$ is the proliferation rate of T cells.}
\label{tab:tab1}
\end{table}
\subsubsection{Average level of virus particles}
\begin{figure}
    \centering
    \includegraphics[width=\linewidth]{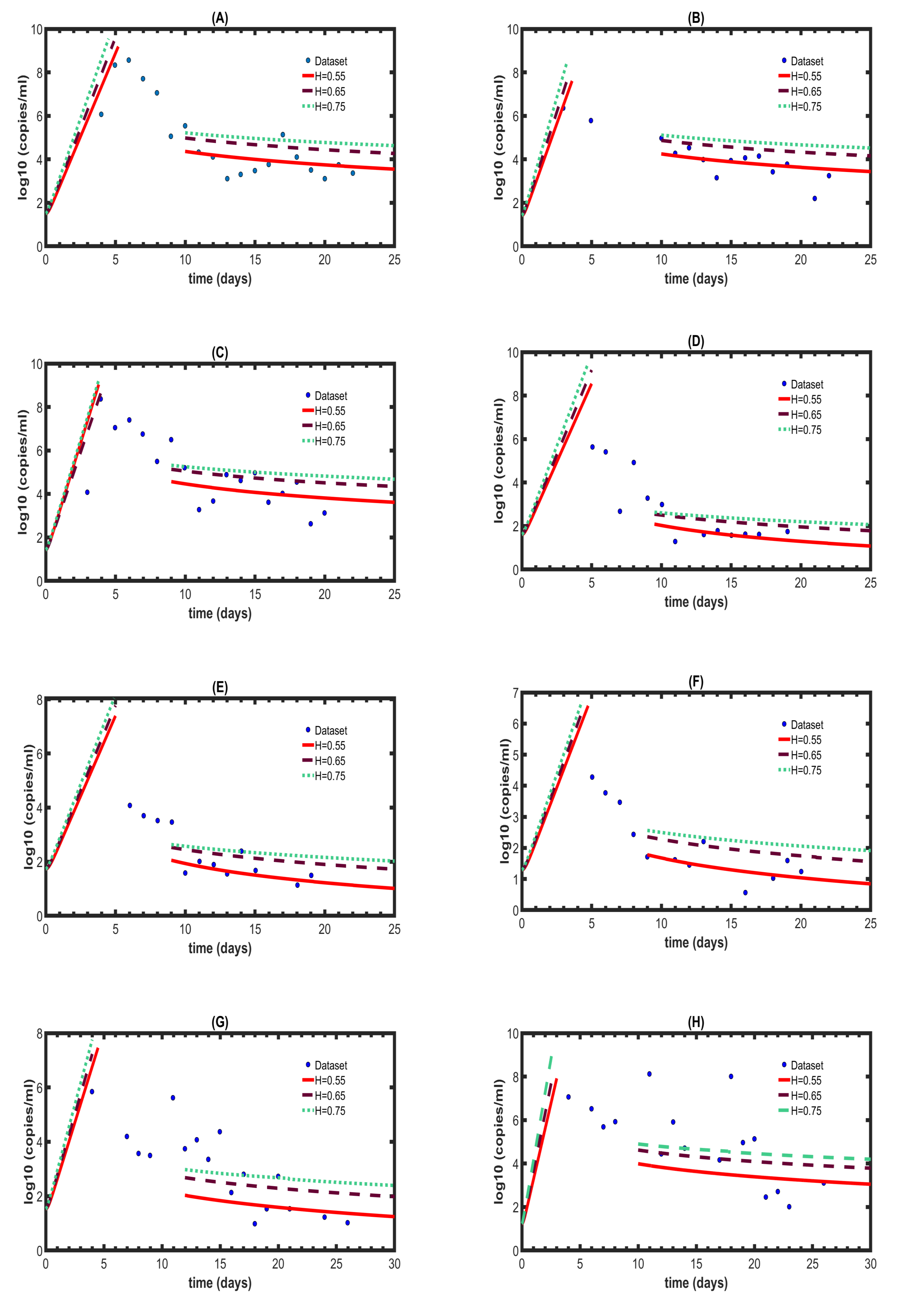}
    \caption{\textbf{Average level of SARS-CoV-2 virus using stochastic immune response model:} 
    The average number of virions at different times is compared for $H$ values $0.55, 0.65, 0.75$. Numerical results from stochastic immune response model (solid lines) are compared to the experimental data (points) of viral load from German patients. Parameters are listed in Table \ref{tab:tab1}.}
    \label{fig4}
\end{figure}

The next step in the analysis of the stochastic immune response model is to compute the average level of virus particles in the system. The analytic expressions for this in different time regimes are obtained using Eq\eqref{prob st} and Eq\eqref{prob lt}. For short times, the average level of virus particles is given by,\\
\begin{equation}\label{av v st}
\left<v(t)\right> = \frac{\sqrt{\vartheta}}{2\sqrt{2 \pi}} \mathbf{E}_{1+b,1}\left(\frac{\pi^{9/2}}{\sqrt 2}\frac{r}{\Gamma(3-b)}\sqrt{\lambda}\sqrt{|(c-p)|} t^{1+b} \right)\\
\end{equation}
where $\vartheta = {\lambda/|(c-p)|}$\\
For long times, this is given by\\
\begin{equation}\label{av v lt}
 \left<v(t)\right> = \frac{2 t^{b-1}}{r a_2 \pi^5 \Gamma{(1-b)}} \mathbf{E}_{b,b}\left(\frac{-2 t^b}{\pi}\frac{|(c-p)|}{\Gamma(3-b)}\right);\\
\end{equation}
After having obtained the expression for mean levels of virus particles, the next step would be to compare it to experimental data. Fig. \ref{fig4} illustrates the comparison between the numerically evaluated average number of virions with that determined clinically in COVID patients from Germany \cite{Wolfel2020}. The rate parameters used in the analysis of each patient are listed in Table \ref{tab:tab1}. The parameter values used have been selected on the basis of fits to data available for experimental viral load in COVID patients in published papers \cite{Wang2020,Hernandez-Vargas2020}. Other parameters have been set to the same values as used in the analysis of the temporal evolution of the joint probability distribution for T cells and virus particles, i.e. in Figs. \ref{fig2} and \ref{fig3}. In short time regime, the initial value $v_0$ is considered to be $100$ copies/ml (which is the lower limit of detection in experiments \cite{Wolfel2020}). In the long time regime, the initial value $v_0$  is in range of $10^{6}-10^{9}$ copies/ml (assuming the peak viral load in patients). 
Eqs. \eqref{av v st} and \eqref{av v lt} are then used to obtain the average virus levels at the early (short time regime) and late stages (long time regime during virus extinction \textit{i.e.} after the viral peak inside the patient's body) of infection. Fig. \ref{fig4} also shows that change in the value of Hurst index, determines the best fit to the experimental data of SARS-CoV-2 load in patients.

The models for $H$ values 0.55, 0.60 and 0.75 have been compared to the SARS-CoV-2 viral load data \cite{Wang2020} from patients. As listed in Table[\ref{tab:tab2}], the model with the best fit has been determined by comparison of the mean square error (MSE) and the Akaike information criterion (AIC) for individual models. The AIC values for the individual models have been calculated using 
\begin{equation}\label{aic}
    \textrm{AIC} = n \log\left(\frac{RSS}{n}\right) + \frac{2 m n}{n-m-1}
\end{equation}
where $n$ is the number of data points, $m$ is the number of unknown parameters and $RSS$ is the residual sum of squares obtained from the fitting routine \cite{Pawelek2012,Burnham2004}. Lower the value of MSE and AIC, better is the model fit to the experimental data of viral load in patients. Fig. \ref{fig4} shows that the maximum viral load or viral peak lies between the predicted results for short time and long time regimes.
 
\begin{table}
\begin{tabular}{ccccc}
    \Xhline{4\arrayrulewidth}
        Patients & & H = 0.55 & H = 0.65 & H = 0.75  \\
    \Xhline{4\arrayrulewidth}
     A & AIC &-8.51 & 4.72 & 12.93\\
      & MSE & 0.48 & 1.17 & 2.03\\
    \hline
     B & AIC&-13.98& 0.04 & 7.45\\
       & MSE& 0.28&0.83 & 1.47\\
    \hline
     C & AIC& 6.36& 7.44 & 13.15\\
       & MSE&1.36& 1.48 & 2.30\\
    \hline
    D& AIC & -3.72&-3.57 & -2.70\\
       & MSE& 0.49&0.50 & 0.55\\
    
    \hline
    E & AIC& -7.37 &-5.78 & -2.52\\
       & MSE& 0.33&0.39 & 0.56\\ 
    \hline
    F & AIC& -8.91 &0.49 & 4.76\\
       & MSE& 0.28&0.79& 1.27\\ 
    \hline
    G & AIC& 24.06 &19.52 & 18.37\\
       & MSE& 5.305&3.74 & 3.42\\
      \hline
    H & AIC& 20.17&16.95 & 16.50\\
       & MSE& 4.40&3.36 & 3.24\\ 
    \Xhline{4\arrayrulewidth}
\end{tabular}
\caption{Comparison of AIC and MSE of three numerically different stochastic immune response models to the experimental data.}
\label{tab:tab2}
\end{table}

\section{Summary and Conclusion}
Understanding the functioning and dynamics of the immune system becomes important given the role that it plays in fighting off infection and disease. However, this becomes non-trivial because just like all the other biological processes at the cellular level, this too shows a lot of variability. Therefore, stochastic models of the immune response, most of which primarily focused on the HIV virus, have been more successful in explaining some of the heterogeneity and variability associated with the system \cite{Hawkins2007,Dalal2008,Wang2019, Fatehi2018,Roy2020}. Earlier theoretical studies that have attempted to explain different aspects of the immune response have been based on deterministic models \cite{Hernandez-Vargas2020, Boianelli2015, Pawelek2012, Best2018, Perelson2002}, which illustrate the mean dynamics but fail to account for the inherent stochasticity and variability of the process. In light of this, in this article, we have developed and analyzed a stochastic version of the immune response model.    
 \\
Our stochastic immune response model is composed of coupled Langevin equations for the T cells and virus particles with two kinds of noise, GWN and fGN, respectively. This allowed us to account for stochasticity within the model itself and obtain the temporal probability distributions of the main variables. 
In this work, we first derived the Fokker-Planck Equation, which we then used to compute the joint probability distribution of T cells and virus particles by making use of the Wilemski-Fixman approximation. This approach allowed us to obtain analytical solutions of the probability distribution functions and the average virus particles in the limit of short and long times, showing how the infection begins and ends (see Figs. \ref{fig2} and \ref{fig3}).
\\
A further advantage of an analytical expression is that a direct comparison can be made between the predicted theoretical dynamics and the experimental results. We have carried out such a comparison with the available SARS-CoV-2 virus data from patients in Germany. At short times, i.e., during the early period of infection, the model predicts that there is a steep rise in the virus levels with time, whereas, at long times, the virus levels drop gradually, in accordance with the model's prediction. As shown in Fig. \ref{fig4}, our Stochastic Immune Response model gives a good fit to the experimental data at both short and long times. 
\\
One of the parameters that is crucial in obtaining good fits is the Hurst index, $H$, which in the case of fGn takes values between 1/2 and 1. $H$ value between 0.5 and 1 corresponds to a system with long-ranged correlated fluctuations and values between 0 and 0.5 stand for anti-correlated time series \cite{lzaro10830}. The H value in the case of fGn, that represents the Non-Markovian viral dynamics, indicates the long-ranged time correlation of the noise $\xi(t)$. 
Higher the H value, greater is the correlation between noises at any time $t$ and the previous time $t'$.  
The $H=1/2$ case on the other hand, is the GWN limit of fGN, where the noise $\xi(t)$ is completely uncorrelated to previous times $t'$ and therefore the two successive times are delta correlated. In addition to the expression of the average level of virions at arbitrary $H$, we have also calculated its expression in the limit of $H=1/2$. In this limit, the expression simplifies to simple exponentials. In the short time regime
\begin{equation}
    \left<v(t)\right> = \frac{\sqrt{\vartheta}}{2\sqrt{2 \pi}}\;\textrm{cosh}\bigg(\bigg(\frac{\pi^{9/2} r a_{1} \sqrt{\vartheta}}{\sqrt{2}}\bigg)^{1/2} t\bigg)
\end{equation}
and in the long time regime it is given by 
\begin{equation}\label{HhalfLt}
    \left<v(t)\right> = \frac{2}{r \pi^5}\frac{|(c-p)|}{ \Gamma{(3-b)}}\;\exp{\bigg(\frac{-2 t |(c-p)|}{\pi}\bigg)}
\end{equation}
In the long time regime, for $H=1/2$, as evident from Eq\eqref{HhalfLt}, there is fast exponential decrease in the virus levels. However, as seen from Fig. \ref{fig4}, the long time regime shows slow temporal decay. This cannot be explained by the GWN limit (Eq\eqref{HhalfLt}) of the average virus levels. Therefore, our model, which includes long ranged noise correlations through fGn provides a more accurate picture of the viral dynamics. For most of the plots in Fig. \ref{fig4}, $H=0.55$ gives a better fit in comparison to other values.
\\
We have also looked at the effect of the strength of the noise on T cell dynamics. The average level of T cells in two different time regimes can be derived using Eq\eqref{prob st} and Eq\eqref{prob lt}. In the short time regime, this is given by 
\begin{equation}\label{av x st}
\left<x(t)\right> = \frac{{1}}{4\sqrt{\pi}}\sqrt{\frac{a}{\gamma}}\: \mathbf{E}_{1+b,1}\left(\frac{\pi^{9/2}}{\sqrt 2}\frac{r}{\Gamma(3-b)}f(\lambda)\sqrt{|(c-p)|} t^{1+b}  \right)\\
\end{equation}
where $f(\lambda)=\sqrt{\lambda}$. At long times, the average level of T cells is given by\\
\begin{equation}\label{av x lt}
 \left<x(t)\right> = \frac{\sqrt{2}}{r \pi^5}\frac{|(c-p)|}{ \Gamma{(3-b)}}\sqrt{\frac{a}{\gamma}}\: t^{b-1} \mathbf{E}_{b,b}\left(\frac{-2 t^b}{\pi}\frac{|(c-p)|}{\Gamma(3-b)}\right);\\
\end{equation} 
From Eq\eqref{av x st}, it is clear that the terms $\sqrt{{a}/ \gamma}$ and $f({\lambda})$ account for the strength of the noise in T cell and viral dynamics respectively. The presence of $f({\lambda})$ in the argument of the Mittag-Leffler function leads to a faster increase in the T cells level with increased $\lambda$ (as clearance rate $\gamma$ of T cells has been fixed). Thus in the short time regime, increase in the strength of the noise will cause an increase in the rate of production of T cells. This phenomenon may affect the system in a way where T cells attain its peak value before the maximum viral load and thus might not be optimized to clear all the virus.
\\
In the long time regime, the expression for the average levels of T cells is a product of a slowly increasing function (power law in time) and a decreasing Mittag-Leffler function. From the expression in Eq\eqref{av x lt}, it is clear that it is only the pre-factor $\sqrt{a/\gamma}$ that accounts for the effect of the strength of noise. Thus, an increase in the value of $a$ will affect the increasing function, but the net effect on T cell levels will not be significant. The T cells dynamics will show long lived dynamics in the long time regime and will have similar values for different strengths of the noise. Thus, our model predicts that the noise in the system may have a major effect at the start of the infection time. The time at which the population of T cells reaches the maximum value within patients is an important factor. The analysis of these levels may then help in determining when the immune response modifiers should be administered to the patients.
\\
The present formulation can also be extended to incorporate increasing complexity by considering the effects of susceptible and infected cells on the immune response system. This model can provide useful insights into the dynamics of various other viral diseases as well, such as measles, influenza, Zika virus, which also have long incubation period as found in SARS-CoV-2 \cite{incubation,zika}. The colored noise incorporated in the model accounts for the long lived dynamics of the virus and can therefore provide more accurate predictions. These stochastic models can therefore help in a better understanding of the immune response system. 

\section*{Data Availability}
The data that supports the findings of this study are available within the article.

\section*{Acknowledgements}
This work is supported by the Science and Engineering Research Board (SERB) MATRICS Grant (Ref. No. MSC/2020/000370) awarded by Department of Science and Technology (DST), India.

\appendix
\section{Derivation of the Fokker-Planck Equation}
We have carried out the derivation of the Fokker-Planck equation by using the methods described in \cite{Chaudhury2006a}\\
The multivariate probability density distribution of $x$ and $v$ at time $t$ is given by
\begin{equation}\label{prob_gen}
     P(x,v,t) = \left<\delta(x-x(t))\delta(v-v(t))\right>
\end{equation}
where $x(t)$ and $v(t)$ are functionals of $\theta(t)$ and $\xi(t)$ respectively.
Differentiation of Eq\eqref{prob_gen} with respect to time $t$ gives
\begin{equation}\label{prob_exp}
\begin{split}
    \frac{\partial}{\partial t}P(x,v,t) = -\frac{\partial}{\partial x}\bigg<{\delta(x-x(t))\delta(v-v(t))\Dot{x}(t)}\bigg>-\frac{\partial}{\partial v}\bigg<{\delta(x-x(t))\delta(v-v(t))\Dot{v}(t)}\bigg>
\end{split}
 \end{equation}
Solution of Eq\eqref{prob_exp} is obtained as follows.
Laplace transform of Eq\eqref{vdyn} provides $\Dot{v}(t)$ such that
\begin{equation} \label{v_laplace}  
    v(t)=v(0)\mathcal{X}(t)+\frac{1}{|(c-p)|}\int_{0}^{t}dt'\xi(t')\mathbf{\phi(t-t')}
\end{equation}
where $\mathcal{X}(t)$ and $\mathbf{\phi(t)}$ are inverse Laplace transforms of 
\begin{equation}
    \mathcal{\Tilde{X}}(s)=\frac{\Tilde{K}(s)}{|(c-p)|+s\Tilde{K}(s)} \quad \text{and} \quad \Tilde{\Phi}(s)= 1-s \mathcal{\Tilde{X}}(s)
\end{equation}
respectively.
By making use of the definition $\mathcal{X}(0)=1$ and eliminating $v(0)$ in Eq\eqref{v_laplace}, we get,
\begin{equation}\label{vdot_fn}
    \Dot{v}(t)=\frac{\mathcal{\Dot{X}}(t)}{\mathcal{X}(t)}v(t)+\frac{1}{|(c-p)|}\mathcal{X}(t)\frac{d}{dt}\left(\int_{0}^{t}dt'\frac{\phi(t-t')\xi(t')}{\mathcal{X}(t)}\right)
\end{equation}
Substituting $\Dot{x}(t)$ from Eq\eqref{xdyn}into Eq\eqref{prob_exp} and taking an average over all realizations of the noise, we obtain,
\begin{equation}\label{prob_exp2}
\begin{split}
    \frac{\partial}{\partial t}P(x,v,t) = \left(-\beta\big(v(t)\big)\frac{\partial}{\partial x} x + \gamma \frac{\partial}{\partial x} x +\frac{1 }{2}a\frac{\partial^2}{\partial x^2}\right)P(x,v,t)-\frac{\partial}{\partial v} \left< \delta(x-x(t))\delta(v-v(t))\Dot{v}(t) \right>
\end{split}
\end{equation}
Substitution of Eq\eqref{vdot_fn} into Eq\eqref{prob_exp2} gives
\begin{equation}\label{prob_int}
\begin{split}
    \frac{\partial}{\partial t}P(x,v,t) = \left(-\beta\big(v(t)\big)\frac{\partial}{\partial x} x + \gamma \frac{\partial}{\partial x} x +\frac{1 }{2}a\frac{\partial^2}{\partial x^2}\right)P(x,v,t) + \eta(t)\frac{\partial}{\partial v} v P(x,v,t) \\-\frac{1}{|(c-p)|}\frac{\partial}{\partial v}\bigg<\delta(x-x(t))\delta(v-v(t))\overline{\xi}(t)\bigg>
\end{split}
\end{equation}
where,
\begin{equation}\label{thetabar}
\begin{split}
  \eta(t) \equiv -\frac{\mathcal{\Dot{X}}(t)}{\mathcal{X}(t)} \quad \textrm{and}\quad \overline{\xi}(t) \equiv \mathcal{X}(t)\frac{d}{dt}\left(\int_{0}^{t}dt'\frac{\phi(t-t')\xi(t')}{\mathcal{X}(t)}\right)
\end{split}
\end{equation}
$\overline{\xi}(t)$ is linearly related to ${\xi}(t)$, which is a Gaussian random function. Therefore, by Novikov’s theorem \cite{Novikov1965}, we get,
\begin{equation}\label{novikov}
\begin{split}
    \left<\delta(x-x(t))\delta(v-v(t))\overline{\xi}(t)\right> = \int_{0}^{t}dt'\left<\overline{\xi}(t)\overline{\xi}(t')\right>\\\left< \frac{\delta}{\delta \overline{\xi}(t')}\delta\left(x-x(t)\right)\delta\left(v-v(t)\right)\right>\\
    = -\frac{\partial}{\partial v}\int_{0}^{t}dt'\left<\overline{\xi}(t)\overline{\xi}(t')\right>\times \left< \delta(x-x(t))\delta(v-v(t)) \frac{\delta v(t)}{\delta \overline{\xi}(t')}\right>
\end{split}
\end{equation}
To find the functional derivative in Eq\eqref{novikov}, Eq\eqref{vdot_fn} is solved by using the integrating factor, such that
\begin{equation}
    v(t) = \exp\left(-\int_{0}^{t}dt'\eta(t')\right) \left[v_0+ \frac{1}{|(c-p)|}\int_{0}^{t}dt'\exp\left( \int_{0}^{t'}dt''\eta(t'') \right)\overline{\xi}(t') \right]
\end{equation}
Here the term $ v_0\exp\left(-\int_{0}^{t}dt'\eta(t')\right)$ is a complementary function which is obtained by satisfying the initial conditions, and the other term is the particular integral. Thus the functional derivative is given by
\begin{equation}\label{fn_derivative}
\begin{split}
    \frac{\delta v(t)}{\delta \overline{\xi}(t')} =\frac{1}{|(c-p)|}\exp\left(-\int_{t'}^{t}dt_1\eta(t_1)\right)
\end{split}
\end{equation}
Therefore,
\begin{equation}\label{fn_derivative2}
\begin{split}
  \left<\delta(x-x(t))\delta(v-v(t))\overline{\xi}(t)\right> = -\frac{\partial}{\partial v}\frac{1}{|(c-p)|}P(x,v,t) \mathcal{D}(t)
\end{split}
\end{equation}
where
\begin{equation}\label{D(t)}
\begin{split}
    \mathcal{D}(t) = \int_{0}^{t}dt'{\overline{\xi}(t)\overline{\xi}(t')}\exp\left(-\int_{t'}^{t}dt_1\eta(t_1)\right)
\end{split}
\end{equation}
Substitution of Eq\eqref{fn_derivative2} and Eq\eqref{D(t)} in Eq\eqref{prob_int}, gives the expression,
\begin{equation}\label{prob_final1}
    \begin{split}
     \frac{\partial}{\partial t}P(x,v,t) = -\beta\big(v(t)\big)\frac{\partial}{\partial x} x P(x,v,t)+\gamma \frac{\partial}{\partial x} x P(x,v,t)+\frac{1 }{2}a\frac{\partial^2}{\partial x^2}P(x,v,t)\\ + \eta(t)\frac{\partial}{\partial v} vP(x,v,t)+ \frac{1}{|(c-p)|^2}\frac{\partial^2}{\partial v^2}P(x,v,t)\mathcal{D}(t)
     \end{split}
\end{equation}
Substituting $\overline{\xi}(t)$ from Eq\eqref{thetabar} into Eq\eqref{D(t)}, we get
\begin{eqnarray}\label{D(t)2}
    \mathcal{D}(t) = \frac{1}{2}\mathcal{X}(t)\mathcal{X}(t') \frac{d}{dt}\frac{1}{\mathcal{X}(t)\mathcal{X}(t')}\int_{0}^{t'}dt_2 \int_{0}^{t}dt_1 \phi(t-t_1)\phi(t-t_2)\xi(t_1)\xi(t_2)
\end{eqnarray}
Solution of Eq [\ref{D(t)2}] is obtained by making use of double Laplace transforms and performing a lengthy algebra as mentioned in \cite{Fox1977a}, which gives
\begin{equation}\label{final D(t)}
    \mathcal{D}(t)=\frac{1}{2}\lambda\; |(c-p)|\; \mathcal{X}^2(t) \frac{d}{dt}\frac{1}{\mathcal{X}^2(t)}\bigg(1-\big(\mathcal{X}(t)\big)^2\bigg)
\end{equation}
Differentiation of Eq\eqref{final D(t)} and substitution of Eq\eqref{thetabar} in Eq\eqref{prob_final1} gives us the desired Fokker-Planck equation.
\begin{equation}
    \begin{split}
    \frac{\partial}{\partial t}P(x,v,t) = \bigg(-\beta\big(v(t)\big)-\beta\big(v(t)\big) x \frac{\partial}{\partial x} + \gamma \frac{\partial}{\partial x} x +\frac{1 }{2}a\frac{\partial^2}{\partial x^2}+ \eta(t)\frac{\partial}{\partial v} v \\+\frac{\lambda}{|(c-p)|}\eta(t)\frac{\partial^2}{\partial v^2} \bigg) P(x,v,t)
    \end{split}
\end{equation}
\section{Derivation of the propagator, $G(x,v,t|x_0,v_0,0)$}
The Green's function follows the equation
\begin{eqnarray}\label{green_def}
    \frac{\partial}{\partial t}G(x,v,t|x_0,v_0,0) = -\textbf{L} G(x,v,t|x_0,v_0,0)
\end{eqnarray}
with the initial condition given by 
\begin{eqnarray}\label{green_delta}
    G(x,v,0|x_0,v_0) = \delta (x-x_0) \delta(v-v_0)
\end{eqnarray}
Hee, the operator \textbf{L} (from Eq\eqref{operator}) is given by 
\begin{eqnarray}\label{operator negative}
        -\textbf{L} = -\beta(v) x \frac{\partial}{\partial x}  +\gamma\frac{\partial}{\partial x}x+\frac{1 }{2}a\frac{\partial^2}{\partial x^2}+\eta(t)\frac{\partial}{\partial v}v+\frac{\lambda}{|(c-p)|}\eta(t)\frac{\partial^2}{\partial v^2}
\end{eqnarray}
where $\beta(v) = rv^m$ ; $m=1$. Green's function can be found explicitly by using the method of Fourier transforms, where,
\begin{eqnarray}\label{fourier G}
    \hat{G}(k_1,k_2,t|x_0,v_0,0)=\frac{1}{2\pi}\int_{-\infty}^{\infty} dk_2 \int_{-\infty}^{\infty} dk_1 \exp(\iota k_1 x) \exp(\iota k_2 v) G(x,v,t|x_0,v_0,0)
\end{eqnarray}
Carrying out the Fourier transform of Eq [\ref{green_def}], we get,
\begin{equation}
\begin{split}
    \frac{\partial}{\partial t} \hat{G}(k_1,k_2,t|x_0,v_0,0)=\bigg(-r \left(\iota \frac{\partial}{\partial k_2}\right) \left( \iota \frac{\partial}{\partial k_1}\right) (\iota k_1) +\gamma + \gamma \left( \iota \frac{\partial}{\partial k_1}\right) (\iota k_1)+\frac{1}{2} a (\iota k_1)^2 +\\\eta(t) + \eta(t) \left(\iota \frac{\partial}{\partial k_2}\right) (\iota k_2) + \frac{\lambda}{|(c-p)|}\eta(t)(\iota {k_2})^2 \bigg)
    \hat{G}(k_1,k_2,t|x_0,v_0,0)
    \end{split}
\end{equation}
Dividing the above equation throughout by $\hat{G}(k_1,k_2,t|x_0,v_0,0)$ gives
\begin{equation}\label{ln G}
\begin{split}
       \frac{\partial}{\partial t}\ln \hat{G}(k_1,k_2,t|x_0,v_0,0) = \iota r k_1 \frac{\partial}{\partial k_1} \frac{\partial}{\partial k_2}\ln \hat{G}(k_1,k_2,t|x_0,v_0)+\gamma - \gamma k_1 \frac{\partial}{\partial k_1 }\ln \hat{G}(k_1,k_2,t|x_0,v_0,0)\\-\frac{1}{2} a k_1^{2}+\eta(t)
        -\eta(t)k_2 \frac{\partial}{\partial k_2}\ln \hat{G}(k_1,k_2,t|x_0,v_0,0)-\frac{\lambda}{|(c-p)|}\eta(t){k_2}^2
\end{split}
\end{equation}
Following the Gaussian ansatz,
\begin{eqnarray}\label{ansatz}
        \ln \hat{G}(k_1,k_2,t|x_0,v_0,0) = \iota k_1 m(t) + \iota k_2 b(t) - \frac{1}{2} {k_1}^2 s(t) - \frac{1}{2} {k_2}^2 y(t) - \frac{1}{2} k_1 k_2 z(t)
\end{eqnarray}
Differentiating Eq\eqref{ansatz} with time and equating terms of corresponding powers of ${k_1}$, ${k_2}$, ${k_1}^2$, ${k_2}^2$ and ${k_1 k_2}$ with those of Eq\eqref{ln G}, we get,
\begin{equation}\label{diff eqn}
\begin{split}
    \frac{d}{dt}m(t)&= -\gamma m(t)-\frac{1}{2}r z(t)\\
    \frac{d}{dt}b(t)&= -\eta (t) b(t)\\
    \frac{d}{dt}s(t)&= -2\gamma s(t)+ a\\
    \frac{d}{dt}y(t)&= -2\eta (t) \left(y(t) - \frac{\lambda}{|(c-p)|} \right)\\
    \frac{d}{dt}z(t)&= -\gamma z(t) + \eta (t) z(t)
    \end{split}
\end{equation}
Using the conditions $m(0)=x_0$, $b(0)=v_0$, $s(0)=0$, $y(0)=0$ and $z(0)=0$, the solution of differential equations in Eq\eqref{diff eqn} turns out to be
\begin{equation}\label{solu gf}
\begin{split}
        m(t) = x_0 e^{-\gamma t};\quad
        b(t) = \mathcal{X}(t) v_0;\quad
        s(t) = \frac{a}{2 \gamma}(1-e^{-2 \gamma t});\\
        y(t) = \frac{\lambda}{|(c-p)|}\bigg( 1- \mathcal{X}^2 (t) \bigg);\quad
        z(t) = 0
\end{split}
\end{equation}
The double inverse Fourier transform of Eq\eqref{ansatz} gives the expression for $G(x,v,t|x_0,v_0,0)$
\begin{align}\label{gf ansatz}
\begin{split}
    G(x,v,t|x_0,v_0,0) = \frac{1}{2 \pi} \frac{2}{\sqrt{4 y(t) s(t) - {z(t)}^2}} \exp\bigg( \frac{-1}{2\left(1-\frac{z(t)^2}{4 s(t) y(t)}\right)} \\\left[\frac{(x-m(t))^2}{s(t)}+ \frac{(v-b(t))^2}{y(t)}- \frac{z(t) (x-m(t))(v-b(t))}{s(t) y(t)} \right]\bigg)
\end{split}
\end{align}
Eq\eqref{gf ansatz} is in the form of a bi-variate Gaussian distribution for two random variables. Substitution of Eq\eqref{solu gf} in Eq\eqref{gf ansatz}, provides the expression for the Green's function mentioned in Eq\eqref{Greens function}, which is,
\begin{equation}\label{Greens function appendix}
\begin{split}
    G(x,v,t|x_0,v_0,0) = \frac{1}{2 \pi}\frac{1}{\sqrt{\frac{a}{2 \gamma}\frac{\lambda}{ |(c-p)|}{(1-e^{-2 \gamma t})}{(1 - \mathcal{X}^2(t))}}}\exp\bigg( \frac{- 1}{2}\bigg( \frac{1}{\frac{a}{2 \gamma}} \frac{(x - x_0 e^{-\gamma t})^2}{ (1-e^{-2 \gamma t})}\\+\frac{1}{\frac{\lambda}{ |(c-p)|} }\frac{(v - v_0 \mathcal{X} (t))^2}{ (1-\mathcal{X}^2 (t))}\bigg) \bigg)
\end{split}
\end{equation}

\bibliography{IR_pf.bib}
\end{document}